# Ultrafast photonic rainbow with controllable orbital angular momentum

Shunlin Huang[1, 2, 3], Peng Wang[1], Xiong Shen[1], Jun Liu[1, 3, *] and Ruxin Li[1, 3]

[1]*State Key Laboratory of High Field Laser Physics and CAS Center for Excellence in Ultra-intense Laser Science, Shanghai Institute of Optics and Fine Mechanics, Chinese Academy of Sciences, Shanghai 201800, China*
[2]*School of Physics Science and Engineering, Tongji University, Shanghai 200092, China*
[3]*University Center of Materials Science and Optoelectronics Engineering, University of Chinese Academy of Sciences, Beijing 100049, China*
[*] *jliu@siom.ac.cn*

**Abstract**
Increasing any degree of freedom of light beam may open a wide application area of this special light beam. Vortex beam with a dimension of orbital angular momentum (OAM) as a useful light source has been widely applied in many fields. Here, unique multicolor concentric ultrafast vortex beams (MUCU-VBs), which are also named ultrafast photonic rainbow, with controllable orbital angular momentum are firstly generated using cascaded four-wave mixing (CFWM) in an yttrium aluminum garnet (YAG) plate. Up to 9 multicolor concentric annular ultrafast vortex sidebands are generated simultaneously. The topological charges of the sidebands, which are controllable by changing the topological charges of the two input pump beams, are measured and in according with the theoretical analysis very well. The novel MUCU-VBs can be manipulated simultaneously in temporal, spatial, spectral domains and OAM state, which open more than one new degree of freedoms of vortex light beam and will be of wide and special applications, such as multicolor pump-probe experiments, simultaneous microparticle manipulation and exploring, and optical communication. Moreover, the special focusing properties of the multicolor ultrafast sidebands, such as multi-focus of different wavelengths, may further extend their application area.

## 1. Introduction

Increasing any degree of freedom of light beam may open a wide application area of this special light beam. Angular momentum as one dimension of light beam including spin angular momentum and orbital angular momentum (OAM), which had been widely studied and applied in many fields in recent years. Comparing to the limited value of spin angular momentum, the value of OAM can own a large number. Light beam with OAM as a useful tool has been widely used in science and technology. Optical vortex beam with phase singularity at the center of the light beam had been proved to be of OAM [1]. Vortex beam has many fascinating properties and special applications [2], such as optical tweezer [3-5] where particles or biomolecules can be manipulated using radiation pressure. Vortex beam had also been used in stimulated emission depletion (STED) microscopy to realize super-resolution [6]. In optical communication, OAM can be used as a new degree of freedom to increase the capacity of optical communication [7, 8].

Numerous methods have been developed to generate vortex beams in the past few decades. Vortex laser beam can be generated directly from a laser cavity using special mode selection technique [9-11]. Mode conversion is another widely used method to obtain vortex beam, where special optical elements such as spiral phase plate [12], Q-plate [6, 13] or spatial light modulator (SLM) [14] are usually used. Furthermore, vortex beam at different wavelength can also be generated by using optical nonlinear process,, such as second-harmonic generation [15, 16], high-harmonic generation [17], and four-wave mixing (FWM) [18-21]. As for a third-order nonlinear process, CFWM had been successfully used to generate multicolor sidebands with a broad spectral bandwidth, as shown in our early works [22-25]. This CFWM can be occurred in nearly any transparent media. By imprinting spiral phases to the pump beams, multicolor sidebands carrying OAM with different topological charges can be obtained [18-21]. Although it is able to generate multicolor sidebands with different OAMs, the sidebands are usually arranged in one row, or the sideband with higher order OAM tends to overlap with the neighbor one. Then it is difficult to collimate all these sidebands simultaneously. Furthermore, due to the noncollinear optical setup, the generated sidebands show deformation in spatial domain, especially for high order sidebands.

Here, we demonstrate the generation of novel multicolor concentric ultrafast vortex beams (MUCU-VBs) in a YAG plate by using CFWM. The generated MUCU-VBs open more than one new degree of freedoms of vortex light beam. In this proof-of-principle experiment, as many as 9 multicolor concentric annular ultrafast vortex sidebands are generated simultaneously. The topological charges (TCs) of the sidebands can be controlled by changing the TCs of the two input pump beams. It means that the novel MUCU-VBs are manipulated simultaneously in temporal, spatial, spectral domains and OAM state. Then they are expected wide potential applications, such as manipulating microparticals or biomolecules at different depths simultaneously, or trapping different micro-objects which are sensitive to different wavelengths at the same time, and also optical communication with increased capacity. In the future, the method can be used to generate MUCU-VBs in other spectral ranges, such as UV range, mid-infrared or far-infrared by replacing the pump beams and the transparent media to the proper ones.

## 2. Experiment

The experimental setup is schematically shown in Fig. 1(a). Ultrafast laser beam from a Ti:sapphire laser system with the central wavelength at 800 nm and repetition rate of 1 kHz is split into two parts using a dichroic mirror. The pulse duration is about 30 fs. The reflective part (beam 1) with wavelength shorter than 800 nm and the transmitted one (beam 2) with wavelength longer than 800 nm. The spectra of the two pump beams are both shown in Fig. 2(b). Beam 1 and beam 2 are both focused by using spherical lenses with the focal lengths of 250 mm and 300 mm, respectively. These two beams are then spatially combined into a 1-mm-thickness YAG plate using the second dichroic mirror. The YAG plate is placed after the focus of beam 1 but before the focus of beam 2. Two spiral phase masks (PM) with $l = 1, 2$ ($2\pi$, $4\pi$ phase circulation) are used to convert the Gaussian pump beams into vortex beams. A delay line is set in the optical path of beam 1 to adjust the time delay between the two pump beams. When the two pump beams are spatiotemporal overlapped in the YAG plate, multicolor sidebands can be generated. The TCs of the sidebands depend on the TCs of the two pump beams. The diameters of the two pump beams in the YAG plate are both about 0.5 mm. The process of the generation of MUCU-VBs is schematically shown in Fig. 1(b). For instance, two rays, line 1 and line 2, from two symmetric spots in one ring of pump beam 1 intersect with another two similar rays, line 3 and line 4 from pump beam 2. The intersection is after the focus of beam 1 and before the focus of beam 2. Although the setup is not a collinear form, symmetric annular multicolor sidebands can be generated, which shows a pseudo-collinear setup. From this pseudo-collinear setup, MUCU-VBs can be generated.

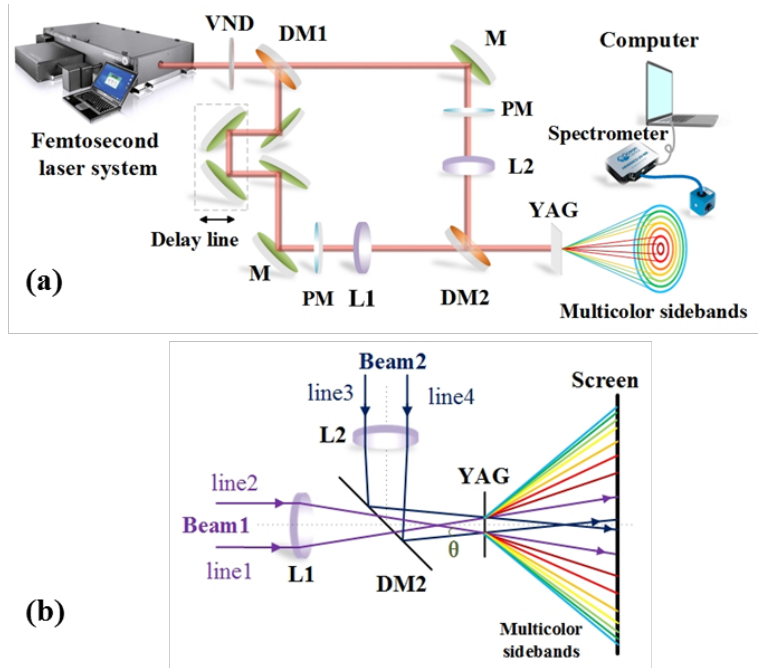

Fig. 1. (a) Experimental setup for generating MUCU-VBs. (b) Schematic for the process of the generation of multicolor sidebands. VND: variable neutral-density filter. DM1(2): dichroic mirror. M: reflective mirror. PM: phase mask. L1(2): plane-convex fused silica lens with the focal length of 250 mm (300 mm). YAG: 1-mm-thickness YAG plate.

## 3. Experimental results and discussion

A typical photograph of the generated multicolor vortex beam on a sheet of white paper placed after the YAG plate is shown in Fig. 2(a). Beautiful multicolor concentric vortex sidebands, which likes a beautiful rainbow, are generated simultaneously. When PM with $l = 1$ or 2 is set in the optical path of beam 1, as many as 9 multicolor sidebands are generated in the experiments, where each ring is one sideband. And the order number of the sidebands increases from inside to outside. All the sidebands are spatially separated and concentric with different diameters. In contrast to conventional vortex beam, of which the diameter of the intensity null dependence on the TC, as it can be seen from Fig. 2(a) and Fig. 1(b) that, the diameters of the generated conical multicolor vortex sideband beams here are independence of the TCs, which can be adjusted by using collimation lenses with different focal lengths. The spectra of the first 6 (5) sidebands when PM $l = 1$ (2) are shown in Fig. 2(b), respectively. It can be seen that all the sidebands are also own broad spectral bandwidths. The spectra of the first 6 concentric sidebands extending from about 550 nm to 750 nm with a spectral bandwidth of about 200 nm. The spectra of the generated sidebands are mainly depended on the crossing angle between the two pump beams. When a $l = 1$ PM and a $l = 2$ PM are used respectively, the spectra of the generated sidebands are overlapped with each other well. It means that the TCs of the sidebands can be conveniently changed by changing the TCs of the pump beams without changing the spectra of the sidebands.

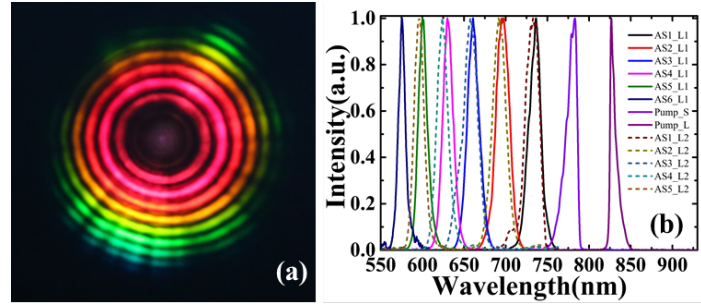

Fig. 2. (a) Photograph of the whole generated multicolor vortex beam on a sheet of white paper placed after the YAG plate (b) The spectra of the first six (five) sidebands when $l = 1$ (2) PM is used in the optical path of beam 1. The two spectra on the right are the spectra of the two pump beams, respectively. The real lines are the spectra when $l = 1$ PM is used and the dash lines are the spectra when $l = 2$ PM is used.

The pulse energy of beam 1 is about 8 times that of beam 2. And the total pump pulse energy here is about 294 μJ. The dependencies of the output pulse energy of the generated multicolor sideband on order number when the total pump pulse energies are 294 μJ, 192 μJ, 149 μJ, respectively, are shown in Fig. 3(a). With the increasing of pump pulse energy, the pulse energy of the generated vortex sideband increased. The first-order vortex sideband has the highest pulse energy of about 1.3 μJ. It means that it is possible to increase the pulse energy of the generated sideband by using higher pump pulse energy. The pulse energy decreased quickly with the increasing of order number for the first 3 or 4 orders. The dependence of the emission angles of the generated vortex sidebands on the order number is also shown in Fig. 3(b). Obviously, the emission angle increased with the increasing of order number. The emission angle is the crossing angle between the sidebands with the central axes of the sidebands themselves.

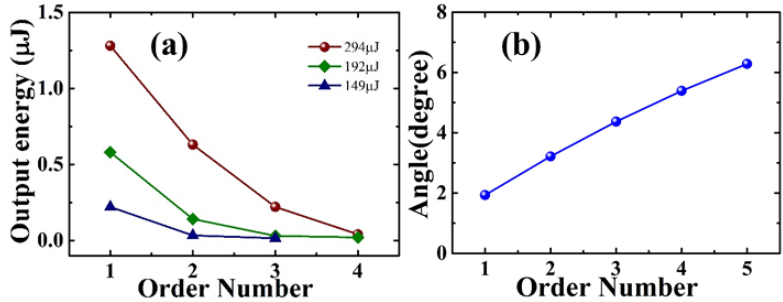

Fig. 3. (a) Dependencies of the output pulse energy of the generated multicolor sideband on order number when total pump pulse energies are 294 μJ, 192 μJ, 149 μJ, respectively. (b) Dependence of the emission angle on order number.

The TCs of the generated multicolor sidebands are measured using interferometry method [26], where the vortex beam interferes with its own mirror image. Different from vortex laser beam from a laser cavity, the generated multicolor sidebands are vortex beams with relative broadband spectra, as shown in Fig. 2(b), hence a Laguerre-Gaussian (LG) field with the spectral bandwidth of about 30 nm is used to simulate the interference pattern of the sideband, which can be written as [26-28]

$$E_1(x,y,w) = (\sqrt{x^2+y^2}/a)^{|l|} \exp[-(x^2+y^2)/a^2] \exp[-(w-w_c)^2/b^2] \exp(il\phi), \quad (1)$$

where $w_c$ is the central frequency of the sideband, $a$ is the waist radius of the LG vortex beam in the space-domain, and $b$ is the waist radius of the Gaussian distribution spectrum of the sideband, $l$ is the TC value, $\phi$ is the azimuthal angle and $\phi=\tan^{-1}(y/x)$. The mirror image of $E_1(x,y,w)$ can be expressed as

$$E_2(x,y,w) = (\sqrt{x^2+y^2}/a)^{|l|} \exp[-(x^2+y^2)/a^2] \exp[-(w-w_c)^2/b^2] \exp(-il\phi), \quad (2)$$

and then $E_1(x,y,w)$ interferes with $E_2(x,y,w)$ with a small angle, the sum of the fields can be calculated as [26]

$$E_3(x,y,w) = E_1(x,y,w)\exp[ikx\sin(\alpha)] + E_2(x,y,w)\exp[-ikx\sin(\alpha)], \quad (3)$$

where $k$ is the wavevector, $\alpha$ is the inclined angle. The interference pattern can be obtained as

$$I_3(x,y,w) = |E_3(x,y,w)|^2. \quad (4)$$

The interference patterns of laser beams with $l = 0, \pm1, \pm2$ with their mirror images are shown in Fig. 4, respectively. The first row shows the experimental results, while the second and third rows show the simulated results. Row 2 shows the isosurfaces of the interference patterns of the broadband laser beams, which are plotted in the $x$-$y$-$w$ coordinates and viewed in $x$-$y$ plane. Row 3 shows the intensity profiles of the interference patterns. The intensity profile is the sum of the interference intensities of all wavelengths and is plotted in the $x$-$y$ plane. The simulated interference patterns for laser beams with different TCs are shown in Fig. 5. As it can be seen that, the interference fringes of the laser beam without vortex, $l = 0$, are parallel to each other. And for $l = m$, the difference of the fringe numbers between the up part and down part of the pattern is $\Delta = 2m$. And the helicity of the vortex beam can be obtained from the forking direction as shown in Fig. 4 for $l = \pm1, \pm2$. The measured interference patterns of the two pump beams with $l = \pm1, \pm2$ are shown in the first row of Fig. 4, respectively. The measured patterns match the simulated patterns well.

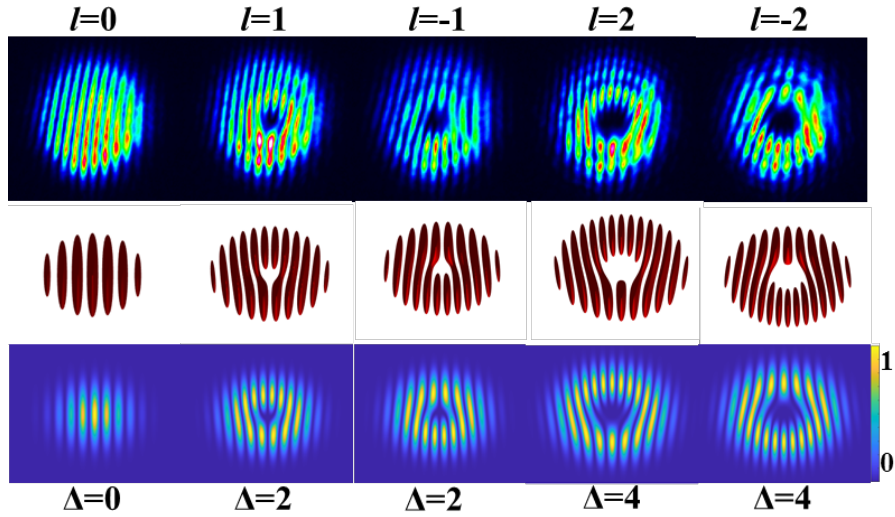

Fig. 4 The interference patterns of the two input laser beams with $l = 0, \pm1, \pm2$. The first row shows the experimental results. And the second and third rows shown the simulated results. Row 2 shows the isosurfaces of the simulated interference patterns of the broadband incident laser beams, which are plotted in the $x$-$y$-$w$ coordinates and viewed in the $x$-$y$ plane. Row 3 shows the intensity profiles of the simulated interference patterns. The intensity profile is the sum of the interference intensities of all wavelengths and is plotted in the $x$-$y$ plane.

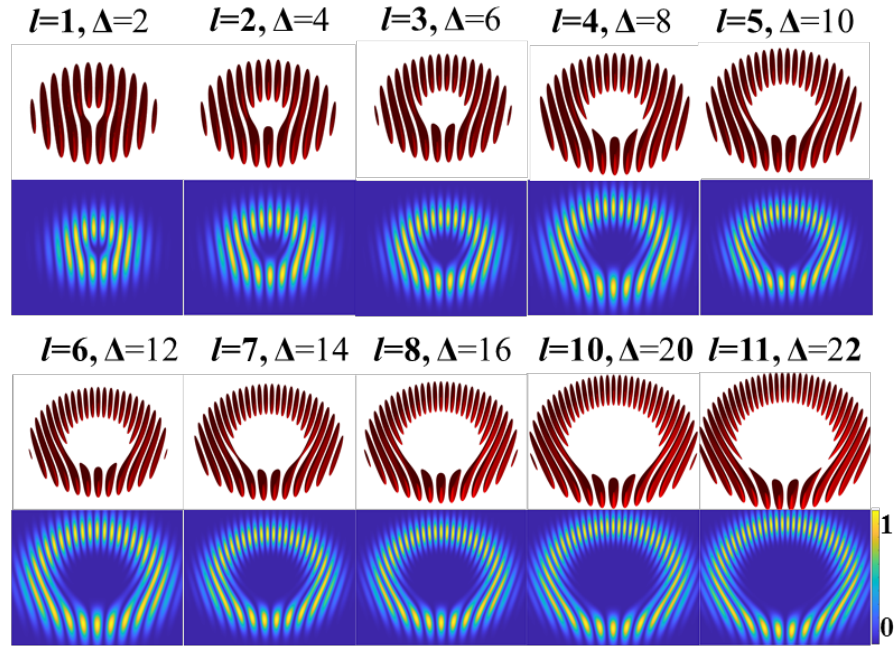

Fig. 5 The simulated interference patterns for laser beams with different TCs. For $l = m$, the difference of the fringe numbers between the up and down parts of the pattern is $\Delta = 2m$, which are noted on the top of each simulated pattern. Row 1 and 3 show the isosurfaces of the simulated interference patterns. Row 2 and 4 show the intensity profiles of the simulated interference patterns.

The measured interference patterns of the generated multicolor sidebands are shown in Fig. 6. Row 1 shows the results when the pump beam with short wavelength, namely beam 1, carrying $l = -1$ and pump beam with long wavelength, beam 2, carrying $l = 0$. Row 2 and row 3 show the results when beam 1 with $l = 2$ or 0, and beam 2 with $l = 0$ or -1, respectively. Row 4 shows the results when both beam 1 and beam 2 are imprinted with spiral phases, where beam 1 with $l = 2$ and beam 2 with $l = -1$. As discussed above, the TCs of the multicolor sidebands can be obtained by calculate the difference of the fringe numbers $\Delta$. And the chirality of the vortex beams can be obtained from the forking directions of the patterns. The values of $\Delta$ and $l$ of each sideband are shown below the pattern. As it is known that, CFWM obeys the energy conservation law and the momentum conservation law as [25] $\omega_{ASm} = \omega_1 + m(\omega_1 - \omega_2)$ and $k_{AS(m-1)} = k_{ASm} - (k_1 - k_2)$ for frequency upshift sidebands. The TC transformation rule is the same as the energy conservation law [19-21], as $l_{ASm} = l_1 + m \cdot \Delta l$, where $\Delta l = l_1 - l_2$. Based on these formulas, the calculated TCs of the first four orders of sidebands in the up three rows are (-2, -3, -4, -5), (4, 6, 8, 10), (1, 2, 3, 4), respectively, while (5, 8, 11) are the TCs of the first three orders of sidebands in the fourth row. The measured TCs and the helicity of the generated sidebands accord well with the calculated ones. The results prove that the generated sidebands carry OAM with different TCs. Moreover, the TCs are controllable by changing the TCs of the two pump beams. The rule of the transfer of TCs during CFWM process is well proved here. Different group of TCs of the pump beams can be used to generate sidebands with different TCs simply. Sideband with TC up to $l = 11$ is obtained and measured clearly based on our experimental conditions, which is the highest measured TC of the sideband generated using CFWM. And it has to note that more than four sidebands were generated in all the four rows, so the actual obtained TC of the sidebands will be larger than 11. The TC will be up to 29 corresponding to the ninth-order sidebands in the fourth row when beam 1 with $l = 2$ and beam 2 with $l = -1$. For higher order sidebands, the number of interference fringes is too large to identify the dense fringes. Moreover, the pules energies of higher sidebands are weak, it is hard to measure their TCs. Sidebands with larger TCs are possible to be generated by using pump beams with larger TCs. As it can be seen that, the interference fringes of the sidebands can be recognized clearly, which shows a good temporal and spatial coherent properties of the generated vortex sidebands. Although the high order sidebands show a little bit ellipse, the fringes can also be seen clearly due to the pseudo-collinear generation setup, from which spatial symmetric vortex sidebands are generated. It shows that the pseudo-collinear generation setup owns high tolerance for generating multicolor sidebands, which shows an advantage of the method used here.

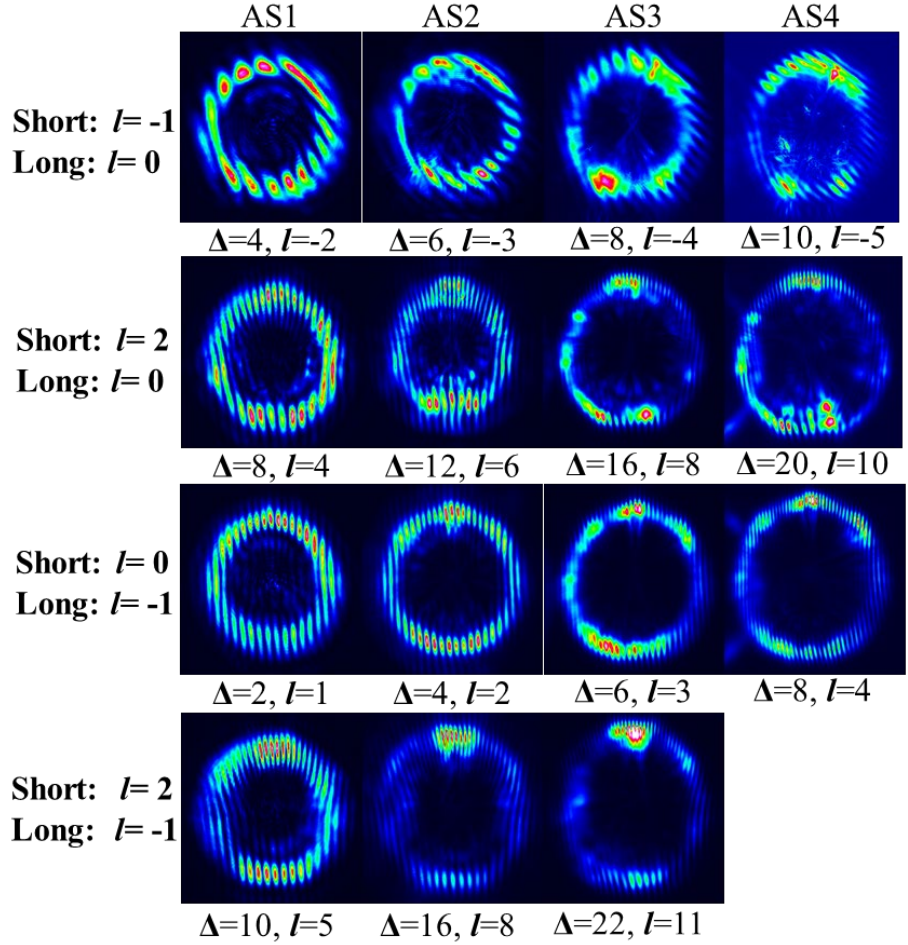

Fig. 6 The measured interference patterns of the generated multicolor sidebands when the two pump beams with different TCs are used. Row 1 shows the results when the short wavelength pump beam, namely beam 1, carrying $l = -1$ and long wavelength pump beam, beam 2, carrying $l = 0$. Row 2 shows the results when beam 1 with $l = 2$ and beam 2 with $l = 0$, row 3 shows the results when beam 1 with $l = 0$ while beam 2 with $l = -1$, row 4 shows the results when beam 1 with $l = 2$ and beam 2 with $l = -1$. Some leaked light inside the rings of the sidebands result from that the lower order sidebands those are not blocked completely.

The focus properties of the generated vortex sidebands are analyzed here. Since vortex beam can be focused into a tiny spot in some applications, the generated MUCU-VBs are collimated using a lens with a focal length of 100 mm and then focused by another lens with a focal length of 500 mm. The focuses of different sidebands can be captured when the CCD is moving from near to far, which means that multi-focus can be formed simultaneously. The focuses can be used to manipulate particles simultaneously at different depths. As an example, the intensity profiles of the focuses of AS1 $l = -2$ and AS2 with $l = -3$ in the first row shown in Fig. 6 are demonstrated in Fig. 7(a) and Fig. 7(b), respectively. The diameter of the focus of AS1 is about 150 μm. The diameter of the focus of AS2 is about 110 μm, which is closed to the diffraction limited of the 500-mm lens, about 85 μm, where the diameter of the input beam is about 10 mm. It can be seen clearly that there is an intensity null at the center of the focus which is a property of vortex beams. High NA focus lens can be used to generate smaller focus. It has to mention that multi-focus is formed simultaneously and each focus can be controlled separately since different focuses are formed by different sidebands. Because the generated multicolor sidebands are spatially separated, any changing of one focus will not affect others. This property may find good application in multi-depths simultaneous microparticle manipulation and exploring, where the optical field is usually affected and changed after the first microparticle or biomolecule is arrested.

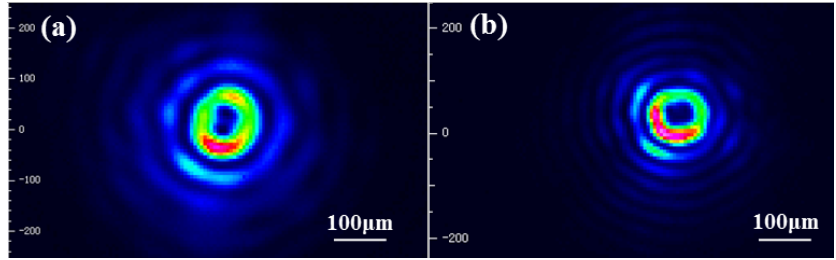

Fig. 7 The intensity profile of the focuses of AS1 with $l = -2$ (a) and AS2 with $l = -3$ (b).

## 4. Conclusion

Novel MUCU-VBs with controllable OAM are firstly generated by using CFWM in a YAG plate. Up to 9 multicolor vortex sidebands can be generated simultaneously. Sideband with TC up to $l = 11$ has been obtained and measured clearly. Sidebands with hundreds TCs are possible to be generated by using pump beams with larger TCs. The properties of the output sidebands, such as the spectra, topological charges, focuses of the sidebands are analyzed. The TCs of the generated sidebands can be controlled simply by changing the TCs of the two pump beams. Furthermore, multi-focus can be formed simultaneously. The multicolor vortex beams can be used simultaneously or separately. The spiral phase mask used here can be replaced by SLM to realize fast modulation of the OAM of the pump beam, and then achieve fast modulation of the TCs of the generated multicolor ultrafast sidebands. The novel MUCU-VBs are manipulated simultaneously in temporal, spatial, spectral domains and OAM state. The generation method can be easily extended to generate MUCU-VBs in other spectral ranges, such as UV range, mid-infrared or far-infrared. The MUCU-VB is absolutely new vortex light source which opens more than one new degree of freedoms of vortex light, which will be of special and wide applications, such as multicolor pump-probe experiments, simultaneous multi-depths microparticle manipulation and exploring, optical tweezer, optical imaging and optical communication.

**Materials and methods**

**Experimental setup**

A 1 kHz Ti:sapphire CPA femtosecond laser system (Legend Elite, Coherent) is used as the laser pulses source. The full spectral bandwidth of the laser source is about 100 nm from 750 nm to 850 nm. Then the laser pulse source is split into two beams with different wavelengths using a dichroic mirror. Pulses with wavelengths longer than 800 nm can pass through the dichroic mirror while pulses with wavelengths shorter than 800 nm will be reflected. This two beams are combined by the second dichroic mirror with transmitted wavelengths shorter than 800 nm and reflective wavelengths larger than 800 nm. Phase masks can be inserted in this two beam paths to modulate the two pump beams into vortex beams with different TCs. The two pump beams are focused into the YAG plate using two conventional spherical focus lenses. The spectra of the pump lights, and the multicolor sidebands are measured using a commercial grating spectrometer (HR4000, Ocean Optics).

**Topological charge measurement**

Interferometry method is used to verify the topological charges of the sidebands due to its high resolution, where the generated sidebands interfere with their own mirror images. In order to produce a vortex beam with opposite helicity and verify the TCs of the sidebands, a home made interferometer is used. The numbers of the reflective mirrors used in the two arms of the interferometer are odd and even, respectively. Hence the TCs of the two interference beams have opposite sign. A charge couple device (CCD) (BC106, Thorlabs) is used to measure the interference patterns of the generated sidebands.

## Acknowledgements

The authors would like to thank Prof. Qiwen Zhan from University of Shanghai for Science and Technology for discussing and reading the paper. This work was funded by the National Natural Science Foundation of China (NSFC) (61527821, 61521093, 61905257, U1930115), Chinese Academy of Sciences (CAS) (the Strategic Priority Research Program (XDB160106)), Shanghai Municipal Natural Science Foundation of China (20ZR1464500).

## Author contributions

S.H. and J.L. conceived and designed the experiment, and drafted the manuscript. S.H. performed all the measurements. P.W. and X.S. involved in the construction of experimental setup and the data analyzed. All authors discussed the results. J.L. and R.L. supervised the project.

## Competing interests

The authors declare no conflicts of interest.

## Data availability

Data underlying the results presented in this paper are not publicly available at this time but may be obtained from the authors upon reasonable request.

## Reference

1. L. Allen, M. W. Beijersbergen, R. J. C. Spreeuw, and J. P. Woerdman, "Orbital angular momentum of light and the transformation of Laguerre-Gaussian laser modes," Phys. Rev. A **45**(11), 8185-8189 (1992).
2. Y. Shen, X. Wang, Z. Xie, C. Min, X. Fu, Q. Liu, M. Gong, and X. Yuan, "Optical vortices 30 years on: OAM manipulation from topological charge to multiple singularities," Light Sci. Appl. **8**, 90 (2019).
3. H. He, M. E. J. Friese, N. R. Heckenberg, and H. Rubinsztein-Dunlop, "Direct observation of transfer of angular momentum to absorptive particles from a laser beam with a phase singularity," Phys. Rev. Lett. **75**(5), 826-829 (1995).
4. L. Paterson, M. P. MacDonald, J. Arlt, W. Sibbett, P. E. Bryant, and K. Dholakia, "Controlled rotation of optically trapped microscopic particles," Science **292**(5518), 912-914 (2001).
5. Y. Yang, Y.-X. Ren, M. Chen, Y. Arita, and C. Rosales-Guzmán, "Optical trapping with structured light: a review," Adv. Photon. **3**(3), 034001 (2021).
6. L. Yan, P. Gregg, E. Karimi, A. Rubano, L. Marrucci, R. Boyd, and S. Ramachandran, "Q-plate enabled spectrally diverse orbital-angular-momentum conversion for stimulated emission depletion microscopy," Optica **2**(10), 900-903 (2015).
7. G. A. Tyler and R. W. Boyd, "Influence of atmospheric turbulence on the propagation of quantum states of light carrying orbital angular momentum," Opt. Lett. **34**(2), 142-144 (2009).
8. I. B. Djordjevic and M. Arabaci, "LDPC-coded orbital angular momentum (OAM) modulation for free-space optical communication. ," Opt. Express **18**(24), 24722-24728 (2010).
9. R. Oron, N. Davidson, A. A. Friesem, and E. Hasman, "Efficient formation of pure helical laser beams," Opt. Commun. **182**(1-3), 205-208 (2000).
10. X. Huang, B. Xu, S. Cui, H. Xu, Z. Cai, and L. Chen, "Direct generation of vortex laser by rotating induced off-axis pumping," IEEE J. Sel. Top. Quantum Electron. **24**(5), 1601606 (2018).
11. N. Li, J. Huang, B. Xu, Y. Cai, J. Lu, L. Zhan, Z. Luo, H. Xu, Z. Cai, and W. Cai, "Direct generation of an ultrafast vortex beam in a CVD-graphene-based passively mode-locked Pr:LiYF4 visible laser," Photonics Res. **7**(11), 1209-1213 (2019).
12. K. Sueda, G. Miyaji, N. Miyanaga, and M. Nakatsuka, "Laguerre-Gaussian beam generated with a multilevel spiral phase plate for high intensity laser pulses," Opt. Express **12**(15), 3548-3553 (2004).

13. E. Karimi, B. Piccirillo, E. Nagali, L. Marrucci, and E. Santamato, "Efficient generation and sorting of orbital angular momentum eigenmodes of light by thermally tuned q-plates," Appl. Phys. Lett. **94**(23), 231124 (2009).
14. N. Matsumoto, T. Ando, T. Inoue, Y. Ohtake, N. Fukuchi, and T. Hara, "Generation of high-quality higher-order Laguerre-Gaussian beams using liquid-crystal-on-silicon spatial light modulators," J. Opt. Soc. Am. A **25**(7), 1642-1651 (2008).
15. K. Dholakia, N. B. Simpson, M. J. Padgett, and L. Allen, "Second-harmonic generation and the orbital angular momentum of light," Phys. Rev. A **54**(5), R3742-R3745 (1996).
16. J. Courtial, K. Dholakia, L. Allen, and M. J. Padgett, "Second-harmonic generation and the conservation of orbital angular momentum with high-order Laguerre-Gaussian modes," Phys. Rev. A **56**(5), 4193-4196 (1997).
17. F. Kong, C. Zhang, F. Bouchard, Z. Li, G. G. Brown, D. H. Ko, T. J. Hammond, L. Arissian, R. W. Boyd, E. Karimi, and P. B. Corkum, "Controlling the orbital angular momentum of high harmonic vortices," Nat. Commun. **8**(1), 14970 (2017).
18. W. Jiang, Q.-f. Chen, Y.-s. Zhang, and G. C. Guo, "Computation of topological charges of optical vortices via nondegenerate four-wave mixing," Phys. Rev. A **74**(4), 043811 (2006).
19. F. Lenzini, S. Residori, F. T. Arecchi, and U. Bortolozzo, "Optical vortex interaction and generation via nonlinear wave mixing," Phys. Rev. A **84**(6), 061801 (2011).
20. J. Strohaber, M. Zhi, A. V. Sokolov, A. A. Kolomenskii, G. G. Paulus, and H. A. Schuessler, "Coherent transfer of optical orbital angular momentum in multi-order Raman sideband generation," Opt. Lett. **37**(16), 3411-3413 (2012).
21. P. Hansinger, G. Maleshkov, I. L. Garanovich, D. V. Skryabin, D. N. Neshev, A. Dreischuh, and G. G. Paulus, "Vortex algebra by multiply cascaded four-wave mixing of femtosecond optical beams," Opt. Express **22**(9), 11079-11089 (2014).
22. J. Liu and T. Kobayashi, "Wavelength-tunable, multicolored femtosecond-laser pulse generation in fused-silica glass," Opt. Lett. **34**(7), 1066-1068 (2009).
23. P. Wang, X. Shen, J. Liu, and R. Li, "Generation of high-energy clean multicolored ultrashort pulses and their application in single-shot temporal contrast measurement," Opt. Express **27**(5), 6536-6548 (2019).
24. S. Huang, P. Wang, X. Shen, and J. Liu, "Multicolor concentric annular ultrafast vector beams," Opt. Express **28**(7), 9435-9444 (2020).
25. J. Liu and T. Kobayashi, "Generation of sub-20-fs multicolor laser pulses using cascaded four-wave mixing with chirped incident pulses," Opt. Lett. **34**(16), 2402-2404 (2009).
26. M. Harris, C. A. Hill, P. R. Tapster, and J. M. Vaughan, "Laser modes with helical wave fronts," Phys. Rev. A **49**(4), 3119-3122 (1994).
27. S. Huang, P. Wang, X. Shen, and J. Liu, "Properties of the generation and propagation of spatiotemporal optical vortices," Opt. Express **29**(17), 26995-27003 (2021).
28. F. He, H. Xu, Y. Cheng, J. Ni, X. Hui, Z. Xu, K. Sugioka, and K. Midorikawa, "Fabrication of microfluidic channels with a circular cross section using spatiotemporally focused femtosecond laser pulses," Opt. Lett. **35**(7), 1106-1108 (2010).